\documentclass[12pt]{article}

\usepackage{amsmath}
\usepackage{amssymb}
\usepackage{amsthm}
\usepackage{mathrsfs}
\usepackage[T1]{fontenc}
\usepackage{enumerate}
\usepackage{color}

\def\eqn{equation}

\def\tfn{transformation}

\def\soln{solution}

\def\sm{sigma model}

\def\dd{Drinfel'd double}
\def\mt{Manin triple}
\def\vbe{beta function equations}
\def\4diml{four-dimensional}
\def\bkg{background}
\def\wrt{with respect to}
\def\-1{^{-1}}
\def\half{\frac{1}{2}}
\def\coor{coordinate}

\def\ca{{\mathfrak a}}
\def\cb{{\mathfrak b}}
\def\cd{{\mathfrak d}}
\def\cg{{\mathfrak g}}
\def\tcg{\tilde{\mathfrak g}}
\def\hcg{\hat{\mathfrak g}}
\def\bcg{\bar{\mathfrak g}}

\def\wt{\tilde}
\def\wh{\widehat}
\def\wwt{\widetilde}
\def\sm{sigma model}
\def\PL{Poisson--Lie }
\def\pltp{Poisson--Lie T-pluralit}

\def\sugra{Generalized Supergravity Equation}

\def\cf{{\mathcal {F}}}
\def\fhat{\widehat \cf}

\newcommand{\unit}{\mathbf{1}}
\newcommand{\nul}{\mathbf{0}}
\newcommand{\A}{\mathscr{A}}

\newcommand{\M}{\mathscr{M}}
\newcommand{\D}{\mathscr{D}}
\newcommand{\G}{\mathscr{G}}
\newcommand{\tG}{\widetilde{\mathscr{G}}}
\newcommand{\hG}{\widehat{\mathscr{G}}}
\newcommand{\bG}{\bar{\mathscr{G}}}

\newcommand{\J}{\mathcal{J}}

\newcommand{\hJ}{\widehat{\mathcal{J}}}

\newcommand{\hd}{\hat{d}}

\newcommand{\N}{\mathscr{N}}

\newcommand{\wb}{\bar}

\title{\PL \tfn s and \sugra s}
\author{Ladislav Hlavat\'y\footnote{hlavaty@fjfi.cvut.cz}
\\ {\em Faculty of Nuclear Sciences and Physical Engineering,}
\\ {\em Czech Technical University in Prague,}
\\ {\em B\v rehov\' a 7, 115 19, Prague 1,}
\\ {\em Czech Republic}
\and
Ivo Petr\footnote{ivo.petr@fit.cvut.cz}
\\ {\em Faculty of Information Technology,}
\\ {\em Czech Technical University in Prague,}
\\ {\em Th\' akurova 9, 160 00, Prague 6,}
\\ {\em Czech Republic}}

\begin{document}
\maketitle

\begin{abstract}
In this paper we investigate \PL \tfn\ of dilaton and vector field $\J$ appearing in \sugra s. While the formulas appearing in literature work well for isometric \sm s, we present examples for which \sugra s are not preserved. Therefore, we suggest modification of these formulas.
\end{abstract}


\tableofcontents


\section{Introduction}

Formula for \PL \tfn\ \cite{klise} of dilaton field accompanying sigma model \bkg\ was given long ago in \cite{unge:pltp}.
Its limitations discussed in \cite{snohla:puzzle} concern the problem of possible appearance of unwanted ''dual'' \coor s of \dd\ in the transformed dilaton.
The problem was partially solved in \cite{dehato,saka2} for special cases where \tfn s of relevant \coor s of \dd\ are linear. 
The price we had to pay was that that in  order to keep invariance \wrt\ \PL \tfn s we had to replace the \vbe\ by \sugra s \cite{Wulff:2016tju} containing not only dilaton, but also additional vector fields for which correct \tfn\ formulas need to be found.
Unfortunately, there are cases for which the \tfn\ of relevant coordinates is not linear or the \PL\ formulas do not provide \soln s of \sugra s.  

In the following we have chosen several examples for which the problem of unwanted \coor s in the transformed dilatons does not appear, and still, it turns out that the original formula \cite{saka2, dehato2, BorWulff:DFT, hlape:pltpbia} for Killing vector field $\J$, which works well for isometric initial \sm s, fails.
These are the cases where the initial \sm s are constructed from \mt\ $(\cd, \cg, \tcg)$ where $\tcg$ is neither Abelian nor unimodular. The purpose of this note is to extend the validity of \PL formulas to these cases. Beside that for NS-NS \sugra s it is not necessary to have formulas for both dilaton and Killing vector $\J$ and one only needs formula for \PL \tfn\ of one-form $X$ that combines these two.
It is given as well.

\section{Basics of \pltp y}\label{basics_of_pltp}

Here we shall recapitulate well known basics of \pltp y with spectators \cite{klise,unge:pltp,hlapevoj} to establish notation.

Sigma models in curved \bkg\ are given by Lagrangian
\begin{equation}\label{Lagrangian}
{\cal L}=\partial_- \phi^{\mu}\cf_{\mu\nu}(\phi)\partial_+ \phi^\nu,\qquad
\phi^\mu=\phi^\mu(\sigma_+,\sigma_-), \qquad \sigma_{\pm} = \tau \pm \sigma
\end{equation}
where map $\phi$ embeds worldsheet $\Sigma\subset \mathbb{R}^2$ into target manifold $\M$ as $\phi:\Sigma \rightarrow \M$. Functions $\phi^{\mu},\ \mu=1,\ldots,\dim \M$, are compositions of $\phi$ and coordinate map on chart of $\M$. Tensor field $\cf=\mathcal G + \mathcal B$ defines metric and torsion potential (Kalb--Ramond field) of the target manifold $\M$.

Assume that there is $d$-dimensional Lie group $\G$ whose action on $\M$ is smooth, proper and free. The action of $\G$ is transitive on its orbits, hence we may locally consider $\M\approx (\M/\G) \times \G = \N \times \G$, $\dim \M = \dim \N + \dim \G = n+d$, and introduce adapted coordinates
\begin{equation}\label{adapted}
\{x^\mu\}=\{s_\alpha,x^a\},\qquad \alpha=1, \ldots,n = \dim \N ,\ \ a=1,
\ldots, d = \dim \G
\end{equation}
where $x^a$ are group coordinates and $s_\alpha$ label the orbits of $\G$. Coordinates $s_\alpha$ are treated as ''spectators'' as they do not participate in \PL\ \tfn s. In this paper we focus on local properties of transformed sigma models and do not impose any conditions following from compactification of $\M$ or $\G$. Discussion of some global issues can be found e.g. in \cite{SomeGlobal} or recent papers \cite{fernandez:Tfolds,hlape:Tfolds}.

\PL\ duality/plurality is based on the possibility to pass between various decompositions of \dd\ $\D$, which is a $2d$-dimensional Lie group whose Lie algebra $\cd$ can be decomposed into double cross sum \cite{majid} of Lie subalgebras $\cg$ and $\tcg$ that are maximally isotropic with respect to non-degenerate symmetric bilinear ad-invariant form $\langle.,.\rangle$. \dd\ with so-called Manin triple $(\cd, \cg, \tcg)$ and Lie groups $\G, \tG$ corresponding to $\cg, \tcg$ is denoted by $\D=(\G|\tG)$. \PL\ dualizable sigma models on $\N \times \G$ are given by tensor field $\cf$ of the form\footnote{Instead of $\cf$ other authors may use its transposition.}
\begin{equation}\label{F}
\cf(s,x)=\mathcal{E}(x)\cdot\left(\unit_{n+d}+E(s) \cdot \Pi(x)\right)^{-1}\cdot
E(s)\cdot \mathcal{E}^T(x)
\end{equation}
where $E(s)$ is spectator-dependent $(n+d)\times (n+d)$ matrix. Denoting generators of $\cg$ and $\tcg$ as $T$ and $\wwt T$, matrix $\Pi(x)$ is given by submatrices $a(x)$ and $b(x)$ of the adjoint representation
$$
ad_{{g}\-1}(\widetilde T) = b(x) \cdot T + a^{-1}(x)\cdot \widetilde T
$$
as
$$\Pi(x)= \left(
\begin{array}{cc}
\nul_n & 0 \\
 0 & b(x) \cdot a^{-1}( x)
\end{array}
\right).$$
Matrix $\mathcal{E}(x)$ reads 
\begin{equation}\label{eRextended}
\mathcal{E}(x)=
\left(
\begin{array}{cc}
 \unit_n & 0 \\
 0 & e(x)
\end{array}
\right)
\end{equation}
where $e(x)$ is $d\times d$ matrix of components of right-invariant Maurer--Cartan form $(dg)g^{-1}$  on $\G$.

For many \dd s several decompositions may exist.
Suppose that we have \sm\ on $\N\times \G$ and the \dd {} splits
into another pair of subgroups $\hG$ and $\bG$. Then we can apply
the full framework of \PL\ T-plurality \cite{klise, unge:pltp} and
find background for sigma model on $\N \times \hG$.

Let Manin triples $(\cd, \cg, \tcg)$ and $(\cd,\hcg, \bcg)$ be two decompositions  of $\cd$ into double cross sum of subalgebras  that are maximally isotropic with respect to $\langle . , . \rangle$. Pairs of mutually dual bases $T_a \in \cg,\
\widetilde{T}^a \in \tcg$ and $\wh T_a \in \hcg,\ \wb{T}^a \in
{\bcg}$, $a=1, \ldots, d,$ 
then must be related by transformation
\begin{equation}\label{C_mat}
\begin{pmatrix}
\wh T \\
\wb T
\end{pmatrix}
 = C \cdot
\begin{pmatrix}
T \\
\widetilde T
\end{pmatrix}
\end{equation}
where $C$ is an invertible $2d\times 2d$ matrix. 
For the following formulas it will be convenient to introduce $d \times d$ matrices $P, Q, R, S$ as
\begin{equation}\label{pqrs}
\begin{pmatrix}
T \\
\widetilde T
\end{pmatrix}
= C^{-1} \cdot
\begin{pmatrix}
\wh T \\
\wb T
\end{pmatrix} =
\begin{pmatrix}
 P & Q \\
 R & S
\end{pmatrix} \cdot
\begin{pmatrix}
\wh T \\
\wb T
\end{pmatrix}
\end{equation}
and extend these to $(n+d)\times (n+d)$ matrices
\begin{equation}
\label{pqrs2}
\nonumber
\mathcal{P} =\begin{pmatrix}\unit_n &0 \\ 0&P \end{pmatrix}, \qquad \mathcal{Q} =\begin{pmatrix}\nul_n&0 \\ 0&Q \end{pmatrix}, \qquad \mathcal{R} =\begin{pmatrix}\nul_n&0 \\ 0&R \end{pmatrix}, \qquad \mathcal{S} =\begin{pmatrix}\unit_n &0 \\ 0& S \end{pmatrix}
\end{equation}
{to accommodate the spectator fields.} 

The sigma model on $\N \times \hG$ obtained from \eqref{F} via \pltp y
is given by tensor field
\begin{equation} \label{Fhat} \widehat{\cf}(s,\hat x)=
\mathcal{\widehat E}(\hat x)\cdot \widehat E(s,\hat x) \cdot
\mathcal{\widehat E}^T(\hat x), \qquad \mathcal{\widehat E}(\hat x)=
\begin{pmatrix}
\unit_n & 0 \\
 0 & \wh e(\hat x)
\end{pmatrix},
\end{equation}
where $\wh e(\hat x)$ is $d\times d$ matrix of components of
right-invariant Maurer--Cartan form $(d\hat g)\hat g^{-1}$  on
$\wh\G$ and
\begin{equation}\label{Fhat2} \wh
E(s,\hat x)=\left(\unit_{n+d}+\wh E(s) \cdot \wh{\Pi}(\hat x)\right)^{-1}\cdot
\wh E(s)  =\left(\wh E\-1(s)+ \wh{\Pi}(\hat x)\right)^{-1}.
\end{equation}
The matrix $\wh E(s)$ is obtained from $E(s)$ in \eqref{F} by formula
\begin{equation}\label{E0hat}
\wh E(s)=(\mathcal{P}+ E(s) \cdot \mathcal{R})^{-1} \cdot
(\mathcal{Q}+E(s) \cdot \mathcal{S}),
\end{equation}
and
$$\widehat\Pi(\hat x)= \left(
\begin{array}{cc}
\nul_n & 0 \\
 0 & \widehat b(\hat x) \cdot \widehat a^{-1}(\hat x)
\end{array}
\right),$$
$$
ad_{{\hat g}\-1}(\wb T) = \widehat b(\hat x) \cdot \wh T + \wh a^{-1}(\hat x)\cdot \wb T.
$$

Conformal invariance up to the first loop requires introduction of dilaton field $\Phi$ satisfying \vbe
\begin{align}
\label{vbetaG}
0 &= R_{\mu\nu}-\frac{1}{4}H_{\mu\rho\sigma}H_{\nu}^{\ \rho\sigma}+2\nabla_{\mu}\nabla_{\nu}\Phi,\\
\label{vbetaB}
0 &= -\frac{1}{2}\nabla^{\rho}H_{\rho\mu\nu}+\nabla^{\rho}\Phi\, H_{\rho\mu\nu},\\
\label{vbetaPhi} 0 &=
R-\frac{1}{12}H_{\rho\sigma\tau}H^{\rho\sigma\tau}+4\nabla_{\mu}\nabla^{\mu}\Phi-4\nabla_{\mu}\Phi\nabla^{\mu}\Phi 
\end{align}
where
$$
H_{\rho\mu\nu}= \partial_\rho \mathcal B_{\mu\nu}+\partial_\mu  \mathcal B_{\nu\rho}+\partial_\nu  \mathcal B_{\rho\mu},
$$
and $\nabla_\mu$ are covariant derivatives \wrt\ metric $\mathcal G$. 

Formula for transformation of dilaton under \pltp y was given
in \cite{unge:pltp} as
\begin{equation}
\wh\Phi(\hat x)= \Phi(y)+\half L(y)-\half\wh L(\hat x)
\label{dualdil}
\end{equation}
where $y$ represent \coor s of group $\G$,
$\Phi(y)$ is the dilaton of the initial model, and terms $L(y)$, $\wh L(\hat x)$ read
\begin{align*}
L(y) &= \ln\Big| \det \left[\left({\bf 1} + \Pi(y) E(s)\right) a(y)\right]\Big|, \\
\wh L(\hat x) &= \ln \Big| \det \left[\left({\bf 1} + \wh\Pi(\hat x) \wh E(s)\right)N\-1\,\wh a(\hat x)\right] \Big|
\end{align*}
with
$$
N=\mathcal{P}^T-\mathcal{R}^T E(s).
$$
The relation between original and new dilaton can be equivalently written as in \cite{dehato2,BorWulff:DFT,Diego} as
\begin{align}
\exp({-2\Phi^{0}(y)}) & :=\exp({-2\Phi(y)})\nonumber\,\frac{(\det \mathcal G(y))^{1/2}}{\det u(y)}\\= \exp({-2\wh \Phi^{0}(\hat x)}) & :=\exp({-2\wh\Phi(\hat x)})\,\frac{(\det \wh{\mathcal{G}}(\hat x))^{1/2}}{\det \hat u(\hat x)}
\label{dilBW}
\end{align}
where $\mathcal{G}$ and $\wh{\mathcal{G}}$ are metrics of \sm s on  $\G$ resp. $\wh \G$ and $u$, $\hat u$ are corresponding matrices of components of left-invariant forms. The fact that $\Phi^{0}(y)=\wh \Phi^{0}(\hat x)$ actually allows one to find relation between original and transformed dilaton. Note that $\Phi^{0}(y)=\Phi(y)+\half L(y)$.

\section{\PL \tfn s and \sugra s}

In the case that the initial dilaton $\Phi(y)$ depends on coordinates $y^k$, we have to express these in terms of $\hat x$ and $\bar x$ to get explicit form of transformed dilaton.
This can be done using relation between two different decompositions of \dd {} elements
\begin{equation}
\label{lgh}
g(y)\wt h(\wt y)=\wh g(\hat x)\bar h(\bar x),\quad 
g\in \G,\ \wt h\in \wt\G,\ \wh g\in \wh\G,\ \bar h\in \bar\G.
\end{equation}
The origin of the puzzle discussed in \cite{snohla:puzzle} is that if
$$
\frac{\partial y^k}{\partial \bar x_j}\neq 0,
$$
formulas \eqref{dualdil}, \eqref{dilBW} give $\wh\Phi$ that may depend not
only on \coor s $\hat x$ of the group $\hG$ but also on \coor s $\bar x$ of $\bG$.

Partial solution of this problem was given in \cite{saka2} for the case of linear dependence
\begin{equation}\label{lineardil0}
 y^k(\hat x,\bar x)= {\hd^k}{}_j \hat x^j+ \bar d^{k j}\bar x_j. 
\end{equation}
It was suggested that in this case we can set $y^k={\hd^k}{}_m \, \hat x^m$ in the formula \eqref{dualdil} and extend
the \vbe\ to (NS-NS) \sugra s \cite{Wulff:2016tju,sugra2}
\begin{align}\label{betaG}
0 &= R_{\mu\nu}-\frac{1}{4}H_{\mu\rho\sigma}H_{\nu}^{\
\rho\sigma}+\nabla_{\mu}X_{\nu}+\nabla_{\nu}X_{\mu},\\ \label{betaB}
0 &=
-\frac{1}{2}\nabla^{\rho}H_{\rho\mu\nu}+X^{\rho}H_{\rho\mu\nu}+\nabla_{\mu}X_{\nu}-\nabla_{\nu}X_{\mu},\\
\label{betaPhi} 0 &=
R-\frac{1}{12}H_{\rho\sigma\tau}H^{\rho\sigma\tau}+4\nabla_{\mu}X^{\mu}-4X_{\mu}X^{\mu}
\end{align}
where
\begin{equation} \label{xform}
X_{\mu}=\partial_{\mu}\Phi + \mathcal J^{\nu} \cf_{\nu\mu},
\end{equation}
$\Phi$ is the dilaton and $ \mathcal J$ is  a vector field.\footnote{The \eqn s \eqref{betaG}-\eqref{betaPhi} without reference to \eqref{xform} were derived in \cite{Mueck} using connection with torsion.} 
For vanishing $ \mathcal J$ the usual \vbe\ are recovered. On the other hand, for \bkg s with $H_{\rho\mu\nu}=0$ equation \eqref{betaB} implies that one can find $\Phi'$ such that $X_{\mu}=\partial_{\mu}\Phi'$. Even though vector field $ \mathcal J$ introduced in \cite{Wulff:2016tju} needs to be Killing field, we do not require this property here as it can be changed by gauge \tfn s keeping $X$ invariant.

Authors of \cite{saka2,hlape:pltpbia} give formula allowing to find components of $\hJ$ for \PL transformed sigma model as
$$
\wh{\mathcal{J}}^\alpha=0,\qquad \alpha=1,\ldots, n=\dim \N,
$$
\begin{equation}\label{kilJ1}
 \mathcal{\wh J}^{\dim \N+ m}(\hat x)=\left(\half{{\bar f}^{ab}}{}_b -\frac{\partial{\Phi^{0}(y)}}{\partial y^k}\Big|_{y=\wh D\cdot \hat x}
 \bar d^{k a}\right){\wh V_a}{}^m(\hat x),
\end{equation}
where $a,b,k,m=  1,\ldots,\dim \G$,
\begin{equation}\label{BarHatD}
    \wh D= \begin{pmatrix} \mathbf{1}_n & 0\\
0 & \hat d \end{pmatrix},
\end{equation}
$\hat d$ is the matrix of coefficients $\hat d^k{}_j$ in \eqref{lineardil0},
$\wh  V_{a}$ are (spectator extended, see \eqref{eRextended}) left-invariant fields of the group
$\hG$, and ${\bar f}^{ba}{}_c$ are
structure constants of the Lie algebra of $\bG$. This formula works well for non-Abelian T-duality, i.e. for \PL \tfn s of \sm s with isotropic \bkg s constructed by \eqref{F}--\eqref{eRextended} where \mt\ is $(\cd, \cg, \ca)$ and $\ca$ is Abelian algebra. We have noticed that 
it does not work for \bkg s obtained  by the \PL \tfn s starting from $(\cd, \cg, \tcg)$ where $\tcg$ is neither Abelian nor unimodular, i.e. when ${{\wt f}^{ab}}{}_b \neq 0$. Examples given below are \bkg s obtained from $(\cd, \ca, \cb)$ where $\cb$ is Bianchi algebra and $\ca$ represents three dimensional Abelian Lie algebra.

From the form of equations \eqref{betaG}--\eqref{betaPhi}  one can see that only one-form $X$ is important for their satisfaction, not
separately dilaton $\Phi$ and vector field $ \mathcal J$. Therefore, for \PL \tfn\ of solutions of \sugra s it would be sufficient to know  only \PL \tfn\ of the one-form $X$ and of the tensor $\cf$. 

Note that $\Phi$ and $ {\mathcal J}$ are not defined uniquely as the form $X$ is invariant \wrt\
gauge \tfn
\begin{equation}\label{gaugetfnX}
   \Phi(x)\mapsto\Phi(x)+\lambda(x),\quad {\mathcal J}^{\kappa}\mapsto{\mathcal J}^\kappa
   -\partial_\nu\lambda\,\cf^{\nu\kappa}
\end{equation}
where $\cf^{\nu\kappa}=(\cf^{-1})^{\nu\kappa}$. This means that we can always choose dilaton vanishing. On the other hand, if $X$ is closed, we can choose $X=d \Phi$ and $\J$ vanishing in which case Generalized Supergravity Equations of Motion become usual \vbe.
Moreover, note that even the form $X$ satisfying equations \eqref{betaG}--\eqref{betaPhi} is not unique. Namely, if $X_\mu$ satisfy the \sugra s, then
\begin{equation}
\label{ambig X} X'_\mu:=X_\mu +\chi_\mu,
\end{equation}
where
\begin{equation}
\label{cond for chi} \nabla_\nu\chi_\mu=0, \quad (X_\mu +2\,\chi_\mu)\chi^\mu=0,
\end{equation}
satisfy the equations as well. Simple example that was mentioned in \cite{hlape:Tfolds} is sigma model given by flat Minkowski metric
\begin{equation}\nonumber
\cf(t, x_2, x_3) = \left(
\begin{array}{cccc}
1 & 0 & 0 & 0 \\
 0 & 0 & 0 & \frac{1}{x_3+1} \\
 0 & 0 & 1 & \frac{x_2}{x_3+1} \\
 0 & \frac{1}{1- x_3} & \frac{x_2}{x_3-1} & \frac{(x_2-2) x_2}{x_3^2-1} \\
\end{array}
\right)
\end{equation}
and
$$
X_\mu=\left(0,0,0,0\right),\qquad \chi_\mu=\left(0,0,0,\frac{1}{1-x_3^2}\right).
$$

In the following we verify whether \bkg s obtained by \pltp y supported by $\wh\Phi$ and $\hJ$ obtained from \eqref{dualdil} and \eqref{kilJ1} satisfy Generalized Supergravity Equations. We give examples where this is not true so that it turns out that formula \eqref{kilJ1} works well for isometric initial \sm s, but not always.
Our original observation was that \eqref{kilJ1} fails to give solution of \sugra s in the case of \PL T-duality when initial \bkg s are constructed from Manin triple $(\cd, \ca, \tcg)$  where $\ca $ is abelian and $\tcg$ is Bianchi 3, 4 or 5 algebra. Surprisingly, the formula \eqref{kilJ1}  does not work even for \PL identity  with these Manin triples, i.e. when the $C$ matrix in \eqref{C_mat} is identity matrix and the algebraic structure is unaffected. Therefore, it is desirable to modify the prescription for vector fields $\mathcal{\wh J}$. Since no changes are necessary for unimodular algebras $\tcg$ in  $(\cd, \cg, \tcg)$, the modified prescription might include terms proportional to ${\widetilde f}^{ab}{}_b$. Gradually searching for more and more involved examples where $\hJ$ is not only constant but depends on spectators or group coordinates we have found formula
\begin{align}
\label{kilJ}\nonumber
 \mathcal{\wh J}^{\dim \N+ m}(\hat x)= &
 \half{{\widetilde f}^{ab}}{}_b \left(\frac{\partial{\widetilde  y_a}}{\partial\bar x_k}{\wh V_k}{}^m(\hat x)-\frac{\partial{\widetilde  y_k}}{\partial\wh x^a} \wh\cf^{km}\right)\\
   & +\left(\half{{\bar f}^{ab}}{}_b -\frac{\partial{\Phi^{0}(y)}}{\partial y^k}\Big|_{y=\wh D\cdot \hat x}
 \frac{\partial{y^k}}{\partial\bar x_a}\right){\wh V_a}{}^m(\hat x)
\end{align}
where 
$\wh  V_{a}$ are left-invariant fields of the 
group $\hG$, ${\widetilde f}^{ba}{}_c$ and ${\bar f}^{ba}{}_c$ are
structure constants of Lie algebras of $\tG,\,\bG$ and $\wh D$ is given by \eqref{BarHatD}. This modification does not change results of \cite{hokico,saka2} and \cite{hlape:pltpbia} because those papers deal with groups for which ${{\widetilde f}^{ab}}{}_b=0$. Finally let us mention that $\wh {\mathcal J}$ obtained from \eqref{kilJ} is not always Killing vector field of $\wh\cf$ (which is not necessary for satisfaction of the NS-NS \sugra s)  but  we can use the gauge \tfn\ \eqref{ambig X} in order that $\hJ$ acquire this property.

Having formulas \eqref{dualdil}, \eqref{xform} and \eqref{kilJ} it is easy to write down prescription for \PL \tfn\ of the form $X$
\begin{align}
\label{plxform}
\wh X_\mu(\hat x) =& \ \frac{\partial{\Phi^{0}(y)}}{\partial y^\nu}\Big|_{y=\wh D\cdot \hat x}\frac{\partial{y^\nu}}{\partial\hat x^\mu}-\half\frac{\partial\wh L(\hat x)}{\partial\hat x^\mu}
   +\hJ^\nu(\hat x)\wh\cf_{\nu\mu}(\hat x) \\
=& \ \left[X_{\nu}(y)-\J^\kappa(y)\cf_{\kappa\nu}(y)\right]\Big|_{y=\wh D\cdot \hat x} \frac{\partial{y^\nu}}{\partial\hat x^\mu}+\half\frac{\partial L(y)}{\partial y^\nu}\Big|_{y=\wh D\cdot \hat x}\frac{\partial{y^\nu}}{\partial\hat x^\mu}  \nonumber 
\\ & -\half\frac{\partial\wh L(\hat x)}{\partial\hat x^\mu}+ \hJ^\nu(\hat x)\wh\cf_{\nu\mu}(\hat x)
\nonumber  
\end{align}
where $\hJ^\nu(\hat x)$ are given by \eqref{kilJ}.
Advantage of the formula \eqref{plxform} is that $\wh X$, differently from $\wh\Phi$ and $\mathcal{\wh J}$, is invariant \wrt\ the gauge \tfn\ \eqref{gaugetfnX}.

\section{Examples}\label{sec:examples}

The difference between formulas \eqref{kilJ1} and \eqref{kilJ} can be shown in examples of \PL \tfn s in \dd s $(\A|\tG)$ where groups $\tG$ are non-semisimple Bianchi groups. We present \tfn s in \dd s denoted\footnote{To denote \dd s and Manin triples we use notation of \cite{snohla:ddoubles} where classification of six-dimensional \dd s was given. The slots in $(.|.)$ denote numbers corresponding to three dimensional algebras $\cg, \tcg$ in Bianchi classification.} as $DD11$ and $DD12$ so the examples are given by \PL pluralities that follow from series of decompositions
\begin{align*}
DD11 &: (1|5) \cong (5|1) \cong ({6_{-1}}|ii5),\\
DD12 &: (1|4) \cong (4|1) \cong ({6_{-1}}|ii4) \cong (ii4|{6_{-1}}).
\end{align*}
Commutation relations defining particular Manin triples and their isomorphisms $C$ are given below. We have also checked pluralities starting with \bkg s on $(2|5), (1|6_{-1})$ and $(6_{-1}|ii5)$ as well as \PL pluralities in 
$$ DD13 : (1|3) \cong (3|1) \cong (3|iii3)$$ obtaining the same results. Beside physical relevance of sigma models on Bianchi groups that often appear in the literature (see e.g. \cite{saka2,hokico} and references therein) the reason for our choice is the fact that in these cases it is possible to solve \eqn s \eqref{lgh} to get the dependence of $y$ and $\wt y$ on $\hat x$ and $\bar x$. In fact, we have $\frac{\partial{\Phi^{0}(y)}}{\partial y^k}=0$ in the presented examples so we need not to solve the equations \eqref{lgh} and check the linear dependence of $y$ on $\hat x$ and $\bar x$. Crucial test of the formula \eqref{kilJ} would be \PL \tfn\ where all terms participate, i.e.$$\frac{\partial{\Phi^{0}(y)}}{\partial y^k}\neq 0,\quad y^k(\hat x,\bar x)= {\hd^k}{}_j \hat x^j+ \bar d^{k j}\bar x_j,\quad {\widetilde f}^{ab}{}_b \neq 0,\quad {\bar f}^{ab}{}_b \neq 0.$$
Unfortunately, it is rather difficult to find such cases.

Group elements will be parametrized as $g=e^{x^1 T_1}e^{x^2 T_2}e^{x^3 T_3}$ where $e^{x^2 T_2}e^{x^3 T_3}$ and $e^{x^3 T_3}$ parametrize their normal subgroups. We deal with \bkg s on four-dimensional manifolds, hence $\dim \N=1$. Spectator $s_1$ is denoted as $t$.

\subsection{\PL\ plurality on \dd\ $(1|5)$} \label{secB5dual}

We shall start our discussion with tensor field\footnote{For typographic reasons we write coordinate indices as subscripts.}
\begin{equation} \label{F15_orig}
\cf (t,y) =\left(
\begin{array}{cccc}
 -1 & 0 & 0 & 0 \\
 0 & \frac{t^2}{t^4+y_2^2+y_3^2} & \frac{y_2}{t^4+y_2^2+y_3^2} &
   \frac{y_3}{t^4+y_2^2+y_3^2} \\
 0 & -\frac{y_2}{t^4+y_2^2+y_3^2} & \frac{t^4+y_3^2}{t^2 \left(t^4+y_2^2+y_3^2\right)} &
   -\frac{y_2 y_3}{t^2 \left(t^4+y_2^2+y_3^2\right)} \\
 0 & -\frac{y_3}{t^4+y_2^2+y_3^2} & -\frac{y_2 y_3}{t^2 \left(t^4+y_2^2+y_3^2\right)} &
   \frac{t^4+y_2^2}{t^2 \left(t^4+y_2^2+y_3^2\right)} \\
\end{array}
\right)
\end{equation}
specifying sigma model on Abelian group $\G$ with corresponding \dd\ $\D=(\G|\tG) = (1|5)$
whose non-trivial commutation relations read
\begin{align}
[\widetilde T^1,\widetilde T^2] & =\widetilde T^2, & [\widetilde T^1,\widetilde T^3] & =\widetilde T^3, & [\widetilde T^1, T_2] & = - T_2, \\ \nonumber
[\widetilde T^1, T_3] & = - T_3, & [\widetilde T^2, T_2] & = T_1, & [\widetilde T^3, T_3] & = T_1.
\end{align}
Background \eqref{F15_orig}, dilaton
\begin{equation}\label{dil15}
\Phi(t,y)=-\frac{1}{2} \ln \left(-t^2 \left(t^4+y_2^2+y_3^2\right)\right)
\end{equation}
and Killing vector $\J= 2\partial_{y_1}$ satisfy \sugra s. 
Corresponding one-form $X$ with components
\begin{equation}
X_\mu(t,y) =\frac{1}{t^4+y_2^2+y_3^2}\left(-\frac{3 t^4+y_2^2+y_3^2}{t},2 t^2,y_2,y_3\right)
\end{equation}
is not closed and we cannot get rid of the vector $\J$ by gauge \tfn s, so reduction of \sugra s to \vbe\ is not possible.

Background \eqref{F15_orig} is actually non-Abelian dual of flat \bkg
\begin{equation} \label{F5_orig}
\fhat (t,\hat x) = \left(
\begin{array}{cccc}
 -1 & 0 & 0 & 0 \\
 0 & t^2 & 0 & 0 \\
 0 & 0 & e^{2 \hat x_1}\, t^2 & 0 \\
 0 & 0 & 0 & e^{2 \hat x_1}\, t^2 \\
\end{array}
\right)
\end{equation}
studied frequently in the literature \cite{GRV,aagl,hokico}. $\fhat$ is invariant \wrt\ action of Bianchi 5 group and dilaton \eqref{dil15} and $\J$ were obtained via \eqref{dualdil} and \eqref{kilJ1}.

Let us note that  $\Phi^{0}(t,y)=-\frac{1}{2} \ln t^3$ and it is not necessary to solve \eqn\ \eqref{lgh} for $y$ to get transformed dilatons $\wh\Phi(\hat x)$. Similar results can be obtained starting from \dd\ $(1|3)$.

\subsubsection{Identity $(1|5) \rightarrow (1|5) $ and full duality $(1|5) \rightarrow (5|1) $}

Let us  check formulas \eqref{dualdil} and \eqref{kilJ1}  applying \PL \tfn\ with $C$ equal to identity matrix to \eqref{F15_orig} and \eqref{dil15}. We recover the original \bkg\ and dilaton, but {\emph{vector field  $ \mathcal{\wh J}={\partial_{y_1}}$  obtained from \eqref{kilJ1} is different from the initial  one and \sugra s are not satisfied}} even in this simple case. Using \eqref{kilJ} instead of \eqref{kilJ1} we get back Killing vector $\hJ=2\partial_{y_1}$  and \sugra s are satisfied. 

By full duality given by
\begin{equation}
C = D_0 :=
\left(
\begin{array}{cccccc}
 0 & \unit_d \\
 \unit_d & 0 \\
\end{array}
\right)
\end{equation}
we get  flat  background \eqref{F5_orig}, but 
dilaton 
\begin{equation} \label{dil51a}
\wh\Phi(t,\hat x)=\frac{1}{2} \ln \left(e^{2 \hat x_1}\right)
\end{equation}
and  vanishing vector $\mathcal{\wh J}$ obtained from formulas  \eqref{dualdil} and \eqref{kilJ1} do not satisfy  \sugra s.

On the other hand, equations \eqref{betaG}--\eqref{betaPhi} are satisfied for  \eqref{F5_orig}, \eqref{dil51a} and
\begin{equation} \label{Kil51a}
\hJ(t,\hat x) =-\frac{1}{t^2}\partial_{\hat x_1}
\end{equation}
that follows from \eqref{kilJ}. Corresponding one-form $\wh X$ vanishes, and by gauge transformation it is possible to eliminate $\wh{\cal J}$ while changing dilaton to
\begin{equation} \label{ddil51a}
\wh\Phi'(t,\hat x)=-c_1 t\,e^{-\hat x_1}+c_2
\end{equation}
with $c_1, c_2$ arbitrary constants. Dilaton \eqref{ddil51a}  and flat  background \eqref{F5_orig} satisfy \vbe.

\subsubsection{Plurality $(1|5) \rightarrow (6_{-1}|ii5) $}

By \PL plurality given by
\begin{equation}
C_{(1|5) \rightarrow (6_{-1}|ii5)}=\left(
\begin{array}{cccccc}
 0 & 0 & 0 & -1 & 0 & 0 \\
 0 & 0 & 0 & 0 & 1 & 0 \\
 0 & 0 & 1 & 0 & 0 & 0 \\
 -1 & 0 & 0 & 0 & 1 & 0 \\
 0 & 1 & 0 & 1 & 0 & 0 \\
 0 & 0 & 0 & 0 & 0 & 1 \\
\end{array}
\right)
\end{equation}
 we get background tensor
\begin{equation} \label{Fii5}
\wh\cf (t,\hat x) =\left(
\begin{array}{cccc}
 -1 & 0 & 0 & 0 \\
 0 & \frac{t^2 e^{2 \hat x_1} \left(\hat x_3^2+1\right)}{t^4+e^{2 \hat x_1}
   \left(\hat x_3^2+1\right)} & \frac{t^4}{t^4+e^{2 \hat x_1}
   \left(\hat x_3^2+1\right)} & \frac{t^2 e^{2 \hat x_1} \hat x_3}{t^4+e^{2
   \hat x_1} \left(\hat x_3^2+1\right)} \\
 0 & -\frac{t^4}{t^4+e^{2 \hat x_1} \left(\hat x_3^2+1\right)} &
   \frac{t^2}{t^4+e^{2 \hat x_1} \left(\hat x_3^2+1\right)} & \frac{e^{2
   \hat x_1} \hat x_3}{t^4+e^{2 \hat x_1} \left(\hat x_3^2+1\right)} \\
 0 & \frac{t^2 e^{2 \hat x_1} \hat x_3}{t^4+e^{2 \hat x_1}
   \left(\hat x_3^2+1\right)} & -\frac{e^{2 \hat x_1} \hat x_3}{t^4+e^{2
   \hat x_1} \left(\hat x_3^2+1\right)} & \frac{e^{2 \hat x_1} \left(t^4+e^{2
   \hat x_1}\right)}{t^2 e^{2 \hat x_1} \left(\hat x_3^2+1\right)+t^6} \\
\end{array}
\right)
\end{equation}
and dilaton
\begin{equation} \label{dilii5}
\wh\Phi(t,\hat x)=-\frac{1}{2} \ln \left(t^2 e^{-2    \hat x_1}\left(t^4+e^{2 \hat x_1} \left(\hat x_3^2+1\right)\right)\right).
\end{equation}
\sugra s are satisfied for
\begin{equation} \label{Kilii5}
\hJ(t,\hat x) =\frac{1}{t^2}\partial_{\hat x_1}
\end{equation}
calculated via \eqref{kilJ}. Vector field $\wh{\cal J}=\partial_{\hat x_2}$ obtained from formula \eqref{kilJ1} does not satisfy \sugra s.

Correct one-form $\wh X$ with components
\begin{equation}
\wh X_\mu(t,\hat x) =\frac{1}{\Delta}\left(-\frac{3 t^4+e^{2 \hat x_1}
   \left(\hat x_3^2+1\right)}{t},e^{2 \hat x_1}
   \left(\hat x_3^2+1\right),2 t^2,e^{2 \hat x_1} \hat x_3\right),
\end{equation}
$$
\Delta=t^4+e^{2 \hat x_1} \left(\hat x_3^2+1\right),
$$
is not closed and \sugra s cannot be reduced to \vbe. Beside that, vector field \eqref{Kilii5} is not Killing of \eqref{Fii5}. However, using the gauge \tfn\ \eqref{ambig X} with $\lambda=\hat x_1$ we get $\wh\Phi'(t,\hat x)=\wh\Phi(t,\hat x) + \hat x_1$ and
$$
\hJ'(t,\hat x) =2\partial_{\hat x_2}
$$
that is Killing vector field of \eqref{Fii5}. One-form $\wh X$ remains unchanged, of course.

\subsection{\PL\ plurality on \dd\ $(1|4)$} \label{secB4}

Next we shall investigate plural sigma models on \dd\ $(1|4)$ with commutation relations
\begin{align}
\label{MT41}
[\wwt T^1,\wwt T^2] & = -\wwt T^2+\wwt T^3, & [\wwt T^1,\wwt T^3] & = -\wwt T^3, & [\wwt T^1, T_2] & = T_2, \\ \nonumber
[\wwt T^1, T_3] & = -T_2+ T_3, & [\wwt T^2, T_2] & = -T_1, & [\wwt T^2, T_3] & = T_1, & [\wwt T^3, T_3] & = -T_1.
\end{align}
Background
\begin{equation} \label{F14_orig}
\cf (t,y) =\left(
\begin{array}{cccc}
 1 & 0 & 0 & 0 \\
 0 & 0 & 0 & \frac{1}{1-y_3} \\
 0 & 0 & 1 & \frac{y_3-y_2}{y_3-1} \\
 0 & \frac{1}{y_3+1} & \frac{y_3-y_2}{y_3+1} & \frac{(y_2-y_3)^2}{y_3^2-1} \\
\end{array}
\right)
\end{equation}
on Abelian group $\G$ was obtained as non-abelian T-dual of flat background
\begin{equation} \label{F4_orig}
\wh\cf (t,\hat x) = \left(
\begin{array}{cccc}
 1 & 0 & 0 & 0 \\
 0 & 0 & e^{-\hat x_1} \hat x_1 & e^{-\hat x_1} \\
 0 & e^{-\hat x_1} \hat x_1 & e^{-2 \hat x_1} & 0 \\
 0 & e^{-\hat x_1} & 0 & 0 \\
\end{array}
\right)
\end{equation}
that is invariant \wrt\ the action of Bianchi 4 group.

Background \eqref{F14_orig}, dilaton
\begin{equation}\label{dill4}
\Phi(t,y)=-\frac{1}{2} \ln \left(1-y_3^2\right)
\end{equation}
and Killing vector $\J=-2\partial_{y_1}$ satisfy \sugra s. Since $\Phi^{0}(y)=0$, we once again do not need to solve \eqref{lgh} to get the transformed dilaton.
 
\subsubsection{Identity $(1|4) \rightarrow (1|4) $}

To check formulas  \eqref{kilJ1} and \eqref{kilJ} we apply \PL \tfn\ with $C$ equal to identity matrix to \eqref{F14_orig} and \eqref{dill4}. We get the original \bkg\ and dilaton. Formula \eqref{kilJ1} gives vector $-\partial_{\hat x_1}$, while from \eqref{kilJ} we obtain Killing vector $\mathcal{\wh J}=-2\partial_{\hat x_1}$. For the former \sugra s are not satisfied, for the latter they hold.

Corresponding one-form
\begin{equation}
\wh X(t,\hat x) = \frac{\hat x_3+2}{\hat x_3^2-1}d \hat x_3
\end{equation}
is closed and we can pass to dilaton 
\begin{equation}
\wh \Phi'(t,\hat x)=\frac{3}{2} \ln (1-\hat x_3)-\frac{1}{2} \ln (\hat x_3+1)
\end{equation}
that together with \eqref{F14_orig} satisfies \vbe.

\subsubsection{Full duality $(1|4) \rightarrow (4|1) $}

By full duality $(1|4) \rightarrow (4|1) $
we get flat background \eqref{F4_orig}, but non-trivial
dilaton
\begin{equation} \label{dil41}
\wh\Phi(t,\hat x)=\frac{1}{2} \ln \left(e^{-2 \hat x_1}\right)
\end{equation}
obtained from \eqref{dualdil} and vanishing vector field $\wh{\cal J}$ obtained from \eqref{kilJ1} do not satisfy  \sugra s. Correct vector field for which these equations are satisfied is 
\begin{equation} \label{Kil41}
\wh{\cal J}(t,\hat x) =e^{\hat x_1}\partial_{\hat x_3}
\end{equation}
and follows from  \eqref{kilJ}. Corresponding  one-form $\wh X$ vanishes and using gauge \tfn\ \eqref{gaugetfnX} we can get $\hJ' = 0$ and dilaton $\wh \Phi' = 0$ satisfying \vbe.

\subsubsection{Plurality $(1|4) \rightarrow (6_{-1}|ii4) $}

Changing the decomposition of \dd\ to $(6_{-1}|ii4)$ using matrix
\begin{equation}
C_{(1|4) \rightarrow (6_{-1}|ii4)}=\left(
\begin{array}{cccccc}
 0 & 0 & 0 & 1 & 0 & 0 \\
 0 & 0 & 0 & 0 & 0 & -2 \\
 0 & 1 & 0 & 0 & 0 & 0 \\
 1 & 0 & 0 & 0 & 0 & -2 \\
 0 & 0 & -\frac{1}{2} & -1 & 0 & 0 \\
 0 & 0 & 0 & 0 & 1 & 0 \\
\end{array}
\right)
\end{equation}
we get background 
\begin{equation} \label{F6m1ii4}
\wh\cf (t,\hat x) =\left(
\begin{array}{cccc}
 1 & 0 & 0 & 0 \\
 0 & -\frac{e^{2 \hat x_1} (\hat x_1+2 \hat x_3)^2}{e^{2 \hat x_1}-4} & -\frac{2}{e^{\hat x_1}-2} & -\frac{e^{2 \hat x_1} (\hat x_1+2
   \hat x_3)}{e^{\hat x_1}-2} \\
 0 & -\frac{2}{e^{\hat x_1}+2} & 0 & 0 \\
 0 & \frac{e^{2 \hat x_1} (\hat x_1+2 \hat x_3)}{e^{\hat x_1}+2} & 0 & e^{2 \hat x_1} \\
\end{array}
\right)
\end{equation} and
dilaton
\begin{equation} \label{dil6m1ii4}
\wh\Phi(t,\hat x)=\frac{1}{2}\ln \left(-\frac{2 e^{2 \hat x_1}}{e^{2 \hat x_1}-4}\right).
\end{equation}
\sugra s are satisfied for \bkg\ \eqref{F6m1ii4}, dilaton \eqref{dil6m1ii4}  and vector field
\begin{equation} \label{Kil6m1ii4}
\wh{\cal J}(t,\hat x)=\left(1-\frac{e^{\hat x_1}}{2}\right)\partial_{\hat x_2}
\end{equation}
obtained from \eqref{kilJ}. One-form $\wh X$ corresponding to  \eqref{dil6m1ii4}  and \eqref{Kil6m1ii4}
\begin{equation}
\wh X(t,\hat x) =\left(\frac{e^{\hat x_1} \left(e^{\hat x_1}-4\right)}{e^{2 \hat x_1}-4}\right)d \hat x_1
\end{equation}
is closed so we can eliminate $\wh{\cal J}$ by gauge transformation.  Dilaton then reads
\begin{equation} \label{ddil6m1ii4}
\wh\Phi'(t,\hat x)=\frac{1}{2} \ln \left(\frac{(2+e^{\hat x_1})^3}{2-e^{\hat x_1}}\right)
\end{equation}
and together with \eqref{F6m1ii4} satisfies \vbe.

\subsubsection{Plurality $(1|4) \rightarrow (ii4|6_{-1}) $}

Plurality given by matrix $C_{(1|4) \rightarrow (ii4|6_{-1})}=D_0\cdot C_{(1|4) \rightarrow (6_{-1}|ii4)}$
gives flat and torsionless background
\begin{equation} \label{Fii46m1}
\wh\cf (t,\hat x) =\left(
\begin{array}{cccc}
 1 & 0 & 0 & 0 \\
 0 & 0 & \frac{3}{1-3 e^{\hat x_2}} & 0 \\
 0 & \frac{1}{e^{\hat x_2}+1} & \frac{3 e^{4 \hat x_2} \hat x_3^2+4 \hat x_1}{2
   e^{\hat x_2}+3 e^{2 \hat x_2}-1} & \frac{e^{\hat x_2} \hat x_2+2 e^{3 \hat x_2}
   \hat x_3}{2 e^{\hat x_2}+2} \\
 0 & 0 & \frac{3 e^{\hat x_2} \hat x_2-6 e^{3 \hat x_2} \hat x_3}{2-6 e^{\hat x_2}} &
   e^{2 \hat x_2} \\
\end{array}
\right)
\end{equation}
and dilaton
\begin{equation} \label{dilii46m1}
\wh\Phi(t,\hat x)=\frac{1}{2} \ln \left(\frac{2 e^{2 \hat x_2}}{-2 e^{\hat x_2}-3 e^{2 \hat x_2}+1}\right).
\end{equation}
\sugra s are satisfied for  \eqref{Fii46m1}, \eqref{dilii46m1} and  Killing vector field
\begin{equation} \label{Kilii46m1}
\wh{\cal J}(t,\hat x) =-\frac{1}{3}\partial_{\hat x_1}.
\end{equation}
One-form $\wh X$ corresponding to  \eqref{dilii46m1} and \eqref{Kilii46m1}
$$
\wh X(t,\hat x) =\frac{2 e^{\hat x_2}}{\left(e^{\hat x_2}+1\right) 
\left(3 e^{\hat x_2}-1\right)} d \hat x_2
$$
is closed and by gauge transformation to dilaton 
\begin{equation} \label{ddilii46m1}
\wh\Phi'(t,\hat x)=\frac{1}{2}\ln\left(\frac{1-3 e^{\hat x_2}}{1+e^{\hat x_2}}\right).
\end{equation}
we get solution of  \vbe.

\section{Conclusions}

It follows from the examples in Sections \ref{secB5dual} and \ref{secB4} and many others  that formulas \eqref{dualdil}  and \eqref{kilJ1} for \PL \tfn s of dilatons and Killing vectors \cite{saka2, dehato2, BorWulff:DFT, hlape:pltpbia} are not universal in the sense that $\wh\Phi$ and $\hJ$ together with transformed \bkg s in general satisfy neither \vbe\ nor \sugra s. They work properly for \tfn s of isometric \sm s based on semi-abelian Manin triples  $(\cd, \cg, \ca)$ but not in other cases.
We propose modification \eqref{kilJ} of formula  \eqref{kilJ1} giving vector fields  $ \mathcal J$ which together with dilatons given by formula \eqref{dualdil}  (when \eqref{lineardil0} holds) satisfy \sugra s for all presented examples (and many others).

From the form of NS-NS sector of Generalized
Supergravity Equations of Motion it is clear that knowledge of  one-form $X$ is important  for their satisfaction, not separately dilaton $\Phi$ and vector field  $ \mathcal J$ that can be changed by gauge \tfn.
Therefore, beside the \PL\ \tfn\ of tensor $\cf$ it is
sufficient to know only the \tfn\ of the form $X$ to keep the \sugra s satisfied. The corresponding formula \eqref{plxform} was checked as well. 

In  many examples the form $X$ is closed so we can  choose ${\mathcal J}$ vanishing by gauge \tfn\ \eqref{gaugetfnX}, and Generalized Supergravity Equations of Motion become usual \vbe. Beside that, the vector fields  $ \mathcal J$ obtained  by the modified  formula \eqref{kilJ} need not be Killing fields of the corresponding \bkg, but the same \tfn\ can be used to restore this property. Resulting dilatons then differ from those obtained from \eqref{dualdil}.

In the language of (modified) Double Field Theory used in \cite{saka2, SUY:gse}, where initial \bkg s are assumed to be isometric, transformation rule \eqref{kilJ1} can be understood as a field redefinition that is necessary for the covariance of DFT equations of motion and the possibility to restore \sugra s. In the future we would like to find whether formula \eqref{kilJ} can be given in terms of DFT as well.


\begin{thebibliography}{99}

\bibitem{klise} C. Klim\v c\'ik and P. \v Severa, \emph{Dual non-Abelian duality and the Drinfeld double}, Phys. Lett. B 351 (1995) 455, [hep-th/9502122].

\bibitem{unge:pltp} R. von Unge, \emph{Poisson--Lie T-plurality}, JHEP 07 (2002) 014, [hep-th/0205245].

\bibitem{snohla:puzzle} L. Hlavat\'y and L. \v Snobl, \emph{Poisson--Lie T--plurality of three-dimensional conformally invariant sigma models II : Nondiagonal metrics and dilaton puzzle}, JHEP 10 (2004) 045, [hep-th/0408126].

\bibitem{dehato} S. Demulder, F. Hassler, and D. C. Thompson, \emph{Doubled aspects of generalised dualities and integrable deformations}, JHEP 02 (2019) 189, [arXiv:1810.11446].

\bibitem{saka2} Y. Sakatani, \emph{Type II DFT solutions from Poisson-Lie T-duality/plurality}, Progress of Theoretical and Experimental Physics 073B04 (2019), [arXiv:1903.12175].

\bibitem{Wulff:2016tju} L.~Wulff and A.~A. Tseytlin, \emph{Kappa-symmetry of superstring sigma model and generalized 10d supergravity equations}, JHEP 06 (2016) 174, [arXiv:1605.04884].

\bibitem{dehato2} S. Demulder, F. Hassler, and D. C. Thompson, \emph{An invitation to Poisson-Lie T-duality in Double Field Theory and its applications}, [arXiv:1904.09992].

\bibitem{BorWulff:DFT} R. Borsato and L. Wulff, \emph{Quantum correction to generalized T-dualities}, Phys. Rev. Lett. 125, (2020) 201603, [arXiv:2007.07902].

\bibitem{hlape:pltpbia} L. Hlavat\'y and I. Petr, \emph{Poisson--Lie plurals of Bianchi cosmologies and Generalized Supergravity Equations}, JHEP 04 (2020) 068, [arXiv:1910.08436].

\bibitem{hlapevoj} L. Hlavat\'y, I. Petr, and V. \v St\v ep\'an, \emph{Poisson--Lie T-plurality with spectators}, J. Math. Phys. 50 (2009) 043504.

\bibitem{SomeGlobal} E. Alvarez, L. Alvarez-Gaume, J. L. F. Barbon, and Y. Lozano, \emph{Some Global Aspects of Duality in String Theory}, Nucl.Phys. B415 (1994) 71, [hep-th/9309039].

\bibitem{fernandez:Tfolds} J. J. Fern\'andez-Melgarejo, J. Sakamoto, Y. Sakatani, and K. Yoshida, \emph{T-folds from Yang-Baxter deformations}, JHEP 12 (2017) 108, [arXiv:1710.06849].

\bibitem{hlape:Tfolds} L. Hlavat\'y and I. Petr, \emph{T-folds as Poisson--Lie plurals}, Eur. Phys. J. C 80, (2020)  892, [arXiv:2004.08387].

\bibitem{majid} S. Majid, \emph{Foundations of quantum group theory}, Cambridge University Press, 1995.

\bibitem{Diego} T. Codina and D. Marques, \emph{Generalized Dualities and Higher Derivatives}, JHEP 10 (2020) 002, [arXiv:2007.09494].

\bibitem{sugra2} G.~Arutyunov, S.~Frolov, B.~Hoare, R.~Roiban, and A.~A.~Tseytlin,  \emph{Scale invariance of the $\eta$-deformed $AdS_5\times S^5$ superstring, T-duality and modified type II equations}, Nucl. Phys. B 903 (2016) 262, [arXiv:1511.05795].

\bibitem{Mueck} W.~M\"uck, \emph{Generalized Supergravity Equations and Generalized Fradkin--Tseytlin Counterterm}, JHEP 05 (2019) 063, [arXiv:1904.06126].

\bibitem{snohla:ddoubles} L. \v{S}nobl and L. Hlavat\'y, \emph{Classification of 6-dimensional real Drinfel'd doubles}, Int. J. Mod. Phys. A 17 (2002) 4043, [math.QA/0202209].

\bibitem{GRV} M. Gasperini, R. Ricci and G. Veneziano, \emph{A problem with non-Abelian duality?}, Phys. Lett. B 319 (1993) 438, [hep-th/9308112].

\bibitem{aagl} E. \'Alvarez, L. \'Alvarez-Gaum\'e, and Y. Lozano, \emph{On non-abelian duality}, Nucl. Phys. B 424 (1994) 155, [hep-th/9403155v4].

\bibitem{hokico} M. Honga, Y. Kima, and E. \'O Colg\'ain, \emph{On non-Abelian T-duality for non-semisimple groups}, Eur. Phys. J. C 78 (2018) 1025, [arXiv:1801.09567].

\bibitem{SUY:gse} Y. Sakatani, S. Uehara, K. Yoshida, \emph{Generalized gravity from modified DFT}, JHEP 1704 (2017) 123 [arXiv:1611.05856].

\end{thebibliography}
\end{document}